\documentclass[letterpaper, 10 pt, conference]{ieeeconf}  

\IEEEoverridecommandlockouts                              

\let\labelindent\relax	

\usepackage{amsmath,amssymb,amsfonts}
\usepackage{mathabx}
\usepackage{algorithmic}
\usepackage{graphicx}
\usepackage{textcomp}
\usepackage[utf8]{inputenc}
\usepackage[T1]{fontenc}
\usepackage{amsmath}
\usepackage{amsfonts}
\usepackage{amssymb}
\usepackage{balance}
\usepackage{graphicx}
\usepackage[rightcaption]{sidecap}
\usepackage{import}
\usepackage{textcomp}
\usepackage[dvipsnames]{xcolor}
\usepackage{bbm}
\usepackage{slashed}
\usepackage{caption}
\usepackage{subcaption}
\usepackage{lipsum}
\usepackage{enumitem}
\usepackage{tabularx}
\usepackage[noadjust]{cite}
\usepackage{easyReview}
\usepackage{bm}
\usepackage{listings}
\usepackage{booktabs}
\usepackage{tikz}
\usetikzlibrary{calc,arrows.meta}
\usepackage{siunitx}
\usepackage{url}
\usepackage{hyperref}

\newcommand{\bbR}{\mathbb{R}}

\usepackage{amsthm}

\theoremstyle{definition}

\theoremstyle{remark}

\title{\LARGE \bf
A Modular Platooning and Vehicle Coordination Simulator\\for Research and Education
}

\author{Kevin Jamsahar, Adrian Wiltz, Maria Charitidou and Dimos V. Dimarogonas
\thanks{This work was supported by the ERC Consolidator Grant LEAFHOUND, the Swedish Research Council, and the Knut and Alice Wallenberg Foundation.}
\thanks{K. Jamsahar, A. Wiltz and D. V. Dimarogonas (corresponding author) are with the Division of Decision and Control Systems, KTH Royal Institute of Technology, SE-100 44 Stockholm, Sweden {\tt\small \{kevinja,wiltz,dimos\}@kth.se}. Maria Charitidou is with the Institute for Systems Research, University of Maryland, College Park, 20742 MD, USA {\tt\small mchar@umd.edu}. Corresponding author: Dimos V. Dimarogonas.}%
}

\begin{document}

\maketitle
\thispagestyle{empty}
\pagestyle{empty}


\begin{abstract}
	This work presents a modular, Python-based simulator that simplifies the evaluation of novel vehicle control and coordination algorithms in complex traffic scenarios while keeping the implementation overhead low. It allows researchers to focus primarily on developing the control and coordination strategies themselves, while the simulator manages the setup of complex road networks, vehicle configuration, execution of the simulation and the generation of video visualizations of the results. It is thereby also well-suited to support control education by allowing instructors to create interactive exercises providing students with direct visual feedback. Thanks to its modular architecture, the simulator remains easily customizable and extensible, lowering the barrier for conducting advanced simulation studies in vehicle and traffic control research. GitHub: \url{https://github.com/KTH-DHSG/Platooning_Simulator}, Youtube: \url{https://www.youtube.com/watch?v=7Ef_6DNhcoE}
\end{abstract}


\section{Introduction}
Recent advances in intelligent transport systems have led to the development of new mobility paradigms that aim to transform transportation at both individual and societal levels. One such paradigm is vehicle platooning. Platooning enables vehicles to travel in close proximity, potentially reducing fuel consumption, CO\textsubscript{2} emissions, and travel time, while increasing traffic throughput and passenger comfort~\cite{Botelho2025}.
Despite these advantages, deploying large fleets of vehicles in real traffic environments remains challenging due to the highly dynamic nature of the environment and the complex interactions among multiple agents. This highlights the need for extensive simulation and testing frameworks that expose vehicles to diverse traffic scenarios and enable systematic evaluation of their robustness, efficiency, and safety under disturbances.

In the context of control, several simulation environments have been designed over the years offering powerful simulation and visualization capabilities for intelligent transportation applications. Among the most widely used  ones  is  CARLA \cite{Dosovitskiy2017}, an open source simulator for high-fidelity 3D simulation of autonomous driving applications. CARLA provides a variety of vehicles, actors and environments for extensive testing of various driving tasks using a rich sensor suite that includes various sensing models for cameras, LiDAR and radars. Although CARLA allows for highly realistic simulations, it does not allow for the testing of distributed control methods limiting its applicability for platoon coordination. SUMO \cite{Lopez2018} on the other hand allows for simulation of large-scale networks of vehicles providing a realistic testbed for message passing and traffic realization allowing to consider traffic lights and various types of road environments. Nevertheless, its architecture does not allow to consider desired dynamic models and control methods prohibiting its immediate use for control and planning. CommonRoad \cite{Althoff2017} allows the user to interactively choose among existing vehicle dynamic models, various types of obstacles and initial conditions. It offers a large set of predefined road scenarios but also gives users the opportunity to design their own road network. Nevertheless, it does not support a  plug-and-play integration of various controllers or vehicle dynamics limiting its flexibility and modularity. 

Motivated by multi-vehicle coordination problems \cite{Frauenfelder2023} and multi-platoon coordination challenges \cite{Charitidou2022b}, which require the simulation of complex nonlinear dynamics and the integration of advanced control methodologies, this work presents an open-source Python-based simulator. The framework enables engineers and practitioners to easily implement and evaluate a wide range of dynamic vehicle models and control strategies on arbitrary road networks with low implementation overhead. More specifically, the simulator supports the definition of various road segments, including straight, curved, and intersection segments, which can be composed into complete road networks. It further allows users to integrate custom vehicle dynamics and control algorithms. A dedicated visualization module enables rendering of simulation results as videos. The proposed architecture is built on modular, interchangeable components, enabling plug-and-play integration of different system elements without requiring additional modifications. As such, it provides a flexible framework for testing and benchmarking complex control and planning algorithms against state-of-the-art approaches. In the context of control engineering education, the simulator enables instructors to create traffic environments in which students can implement and test their own control algorithms, while receiving immediate visual feedback.

\begin{figure}[b]
	\vspace{-\baselineskip}
	\centering
	\includegraphics[width=1.0\columnwidth]{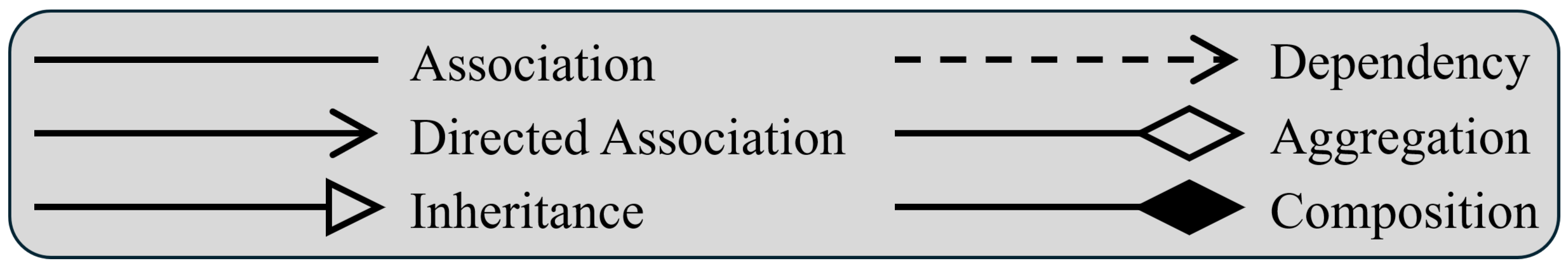}
	\caption{Recap -- Associations in UML}
	\label{fig:UMLnotations}
\end{figure}

\section{Inside the Simulator - The Modular Structure}
\label{sec:modular simulator}

In this section, we present our Python-based simulator and its features. To formally illustrate its modular structure, we employ the \emph{Unified Modeling Language}~\cite{omg2017,Fowler2004} as an established formalism. An overview of the most important associations in UML is provided in Fig.~\ref{fig:UMLnotations}. 

\subsection{Overview of System Modules}
\begin{figure}[t]
	\centering
	\includegraphics[width=0.7\columnwidth]{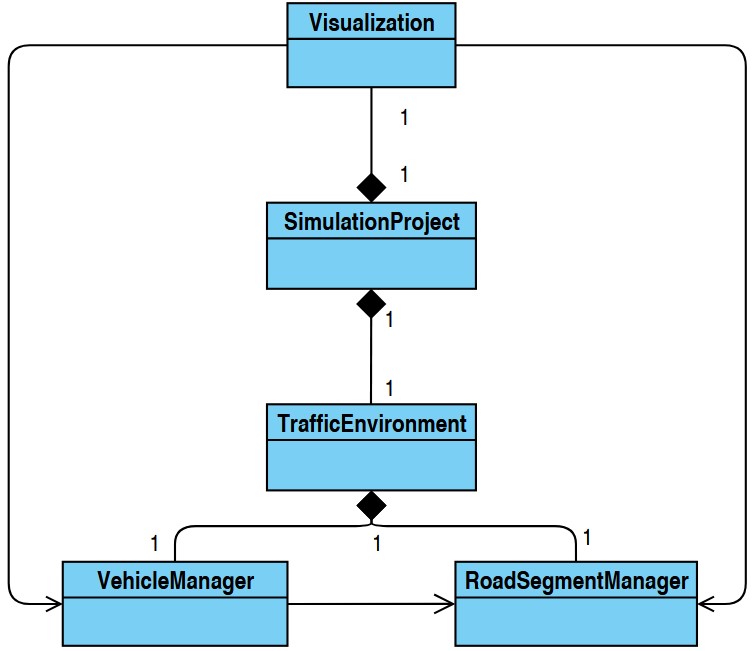}
	\caption{Class diagram of the simulator's core components.}
	\label{fig:overview}
\end{figure}

The \texttt{SimulationProject} class is the entry point for running simulations. It instantiates the \texttt{TrafficEnviron}-\texttt{ment} and manages the simulation loop by incrementing the simulation time, updating vehicles and roads at each time step, logging data for later analysis and recording a video through the \texttt{Visualization} class. The actual simulation of the vehicles in the road network is carried out by the \texttt{TrafficEnvironment}, which acts as the central integration layer for two of the most important managing classes: the \texttt{RoadSegmentManager}, which allows for the construction of the road network by composing it out of standardized road segments and managing it during simulation, and the \texttt{VehicleManager} for configuring and simulating the individual vehicles. The interdependencies of the classes are shown in Fig.~\ref{fig:overview}. Both manager classes bundle functionality provided through supporting modules presented next.

\subsection{Road Segment Manager}

\begin{figure}
	\centering
	\includegraphics[width=0.95\linewidth]{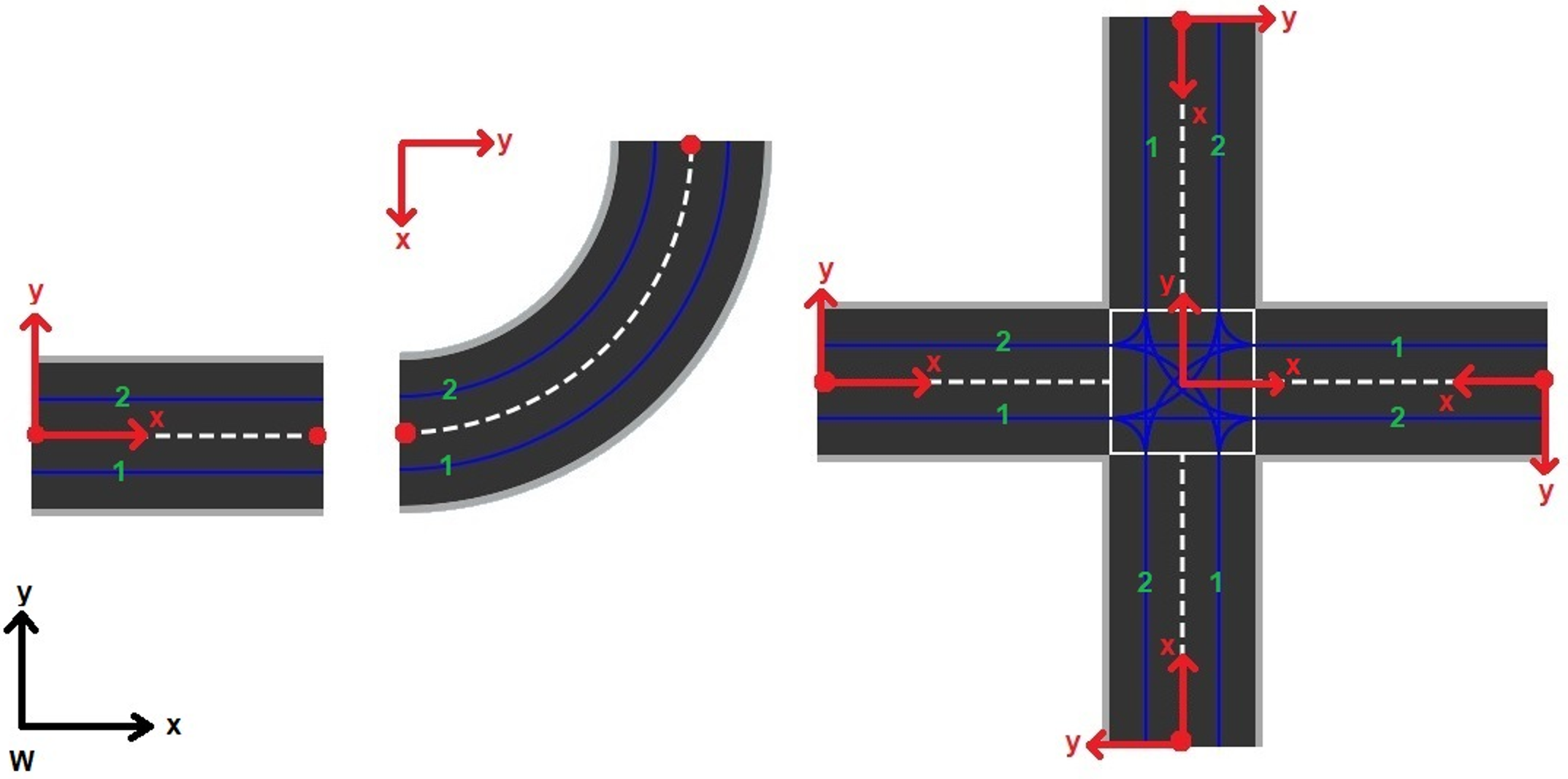}
	\caption{(a)~Straight, (b)~curved, and (c)~intersection segment with respective connection points (red dots). Lanes are numbered as indicated, and the blue lines mark the lane centers.}
	\label{fig:connection_points}
	\vspace{-\baselineskip}
\end{figure}

The \texttt{RoadSegmentManager} provides road segments for straight roads, curves and intersections as shown in Fig.~\ref{fig:connection_points}, which can be assembled blockwise to a road network using functions \texttt{create\_road\_segment()} and \texttt{connect\_road\_segments()}. While running the simulation, the manager supplies the vehicles in the \texttt{VehicleManager} with information about the road network, allowing to imitate the perception of the road through their sensors. Moreover, the vehicles and their control algorithms can query, if required, the \texttt{RoadSegmentManager} to determine speed limits, lane boundaries and lane width, or access the center of the current lane in form of a trajectory. The \texttt{RoadSegmentManager} is complemented by a \texttt{VirtualParkingLot} receiving those vehicles that leave the road network at open ends and feeding them back into the simulation according to certain patterns as detailed further below. Another function allows for the automatic generation of road segments connecting two user specified open ends in the road network, thereby simplifying its setup. In the following, we elaborate on each of these features.

\begin{figure}
	\centering
	\includegraphics[width=0.8\linewidth]{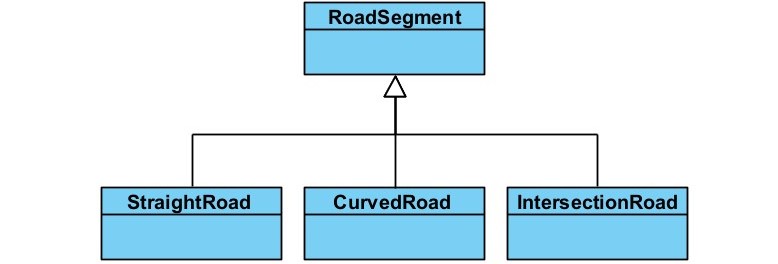}
	\caption{Class diagram illustrating the relationship between the abstract RoadSegment superclass and its subclasses.}
	\label{fig:road_segment_classes}
	\vspace{-\baselineskip}
\end{figure}

\subsubsection{Road Segments} 
\begin{figure}
	\centering
	\includegraphics[width=1.0\columnwidth]{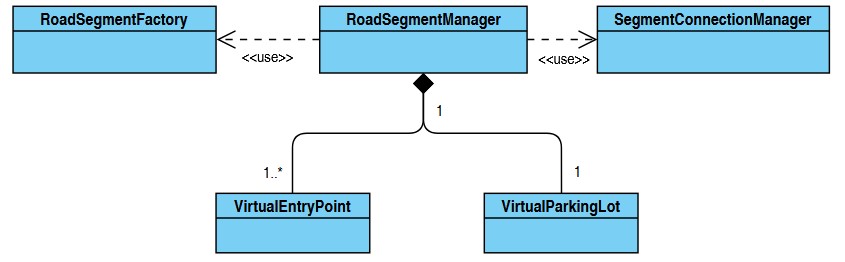}
	\caption{Class diagram \texttt{RoadSegmentManager}.}
	\label{fig:road_segment_manager_architecture}
	\vspace{-\baselineskip}
\end{figure}

The various road segments are represented by the classes \texttt{StraightRoad}, \texttt{CurvedRoad} and \texttt{IntersectionRoad}, each implementing the \texttt{RoadSegment} interface allowing for a unified handling through the \texttt{RoadSegmentManager}, see Fig.~\ref{fig:road_segment_classes}. While the overall road network is represented in the global world coordinate frame $ W $, each of the road segments possesses at least one local coordinate frame as shown in Fig.~\ref{fig:connection_points}. The straight and the curved road segments possess one local coordinate system each, whereas the intersection segment is composed out of four straight subsegments and the intersection itself, motivating the definition of more than one local coordinate system. Each road segment is defined by a \texttt{segment\_id}, its \texttt{length} and \texttt{orientation}, the number of \texttt{lanes}, the \texttt{lane\_width} and the placement of its \texttt{local\_origin} in global coordinates. By convention, the lanes are numbered starting from $ 1 $ and increase as indicated in Fig.~\ref{fig:connection_points}. The center of each lane, indicated by the blue lines, is defined as a trajectory and can be queried by the vehicles as reference trajectory (see below). In order to construct a road network from the separate road segments, the segments must be connected via their \texttt{ConnectionPoint}s, illustrated by the red dots in Fig.~\ref{fig:connection_points}. Two road segments can be connected to each other if they are compatible, meaning that their number of \texttt{lanes} and \texttt{lane\_width} coincide. In that case, any road segment can be connected at a user-specified~\texttt{ConnectionPoint} to one of the \texttt{ConnectionPoint}s of another compatible road segment through the function \texttt{connect\_road\_segments()}. The alignment of the two segments is automatically handled by updating the \texttt{local\_origin} and \texttt{orientation} of the segment to be connected. To preserve consistency and avoid ambiguity, any \texttt{ConnectionPoint} can be connected to at most one other \texttt{ConnectionPoint}, which is internally checked upon connection. The functions for instantiating and connecting road segments are implemented in the classes \texttt{RoadSegmentFactory} and \texttt{SegmentConnectionManager}, see Fig.~\ref{fig:road_segment_manager_architecture}.

\subsubsection{Automatic Road Generation} 
\begin{figure}
	\centering
	\includegraphics[width=1.0\columnwidth]{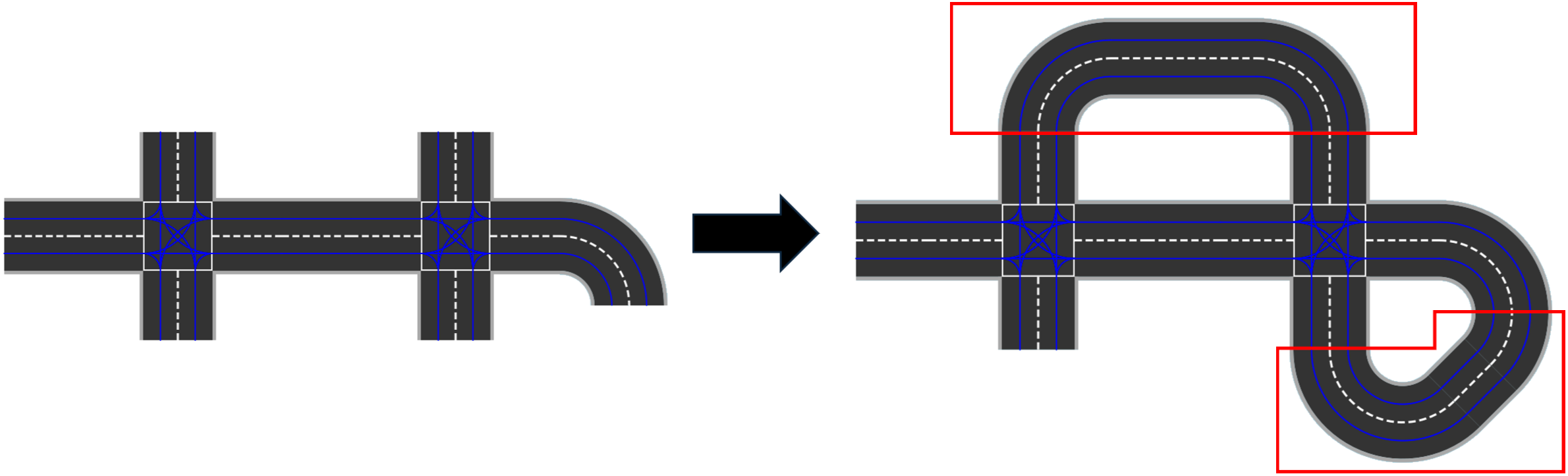}
	\caption{Automatic generation of road segments to complement the road network by using \texttt{create\_connection()}.}
	\label{fig:road_connection}
	\vspace{-\baselineskip}
\end{figure}

To facilitate the creation of road networks, the simulator provides an option to automatically generate road segments that connect two open ends in the existing road network with each other, see Fig.~\ref{fig:road_connection}. To this end, the user selects two road segments, specifies an unconnected \texttt{ConnectionPoint} on each, and applies the \texttt{create\_connection()} method of the \texttt{RoadSegmentManager}; if any of the \texttt{ConnectionPoint}s is already  connected, an error is raised. The automatic road generation is based on Dubin's path~\cite{Dubins1957} and generates a curve-straight-curve (CSC) sequence of road segments connecting the two selected \texttt{ConnectionPoint}s. In degenerate cases, where one or more of the road segments in the sequence would take a vanishing size, not all of these road segments are practically instantiated. For a more detailed discussion of the connection algorithm, we refer to 
~\cite{Jamsahar2025}. 

\subsubsection{Virtual Parking Lot and Re-entry Points} For handling all vehicles that leave the road network through one of its open ends, the simulator provides the possibility to add a \texttt{VirtualParkingLot} using the method \texttt{create\_virtual\_parking\_lot()} of the \texttt{RoadSegmentManager}. The parking lot does not act as a road segment itself, but is a purely virtual construct that collects the vehicles leaving the road network, collecting them into platoons of a fixed, user-specified length (\texttt{platoon\_size}), and releases them once complete with a normally distributed departure time and given mean (\texttt{time\_mean}) and variance (\texttt{time\_variance}) to the network. The platoon enters the road network through an entry point randomly chosen from a list defined as \texttt{exit\_points} of the virtual parking lot. Each entry in \texttt{exit\_points} is a tuple of a road segment-id and a \texttt{ConnectionPoint} of the respective segment. Of course, by setting the \texttt{platoon\_size} to one, the road network can be also configured such that all vehicles in the \texttt{VirtualParkingLot} randomly re-enter the road-network individually through any of the configured re-entry points.

\begin{figure}
	\centering
	\includegraphics[width=0.6\columnwidth]{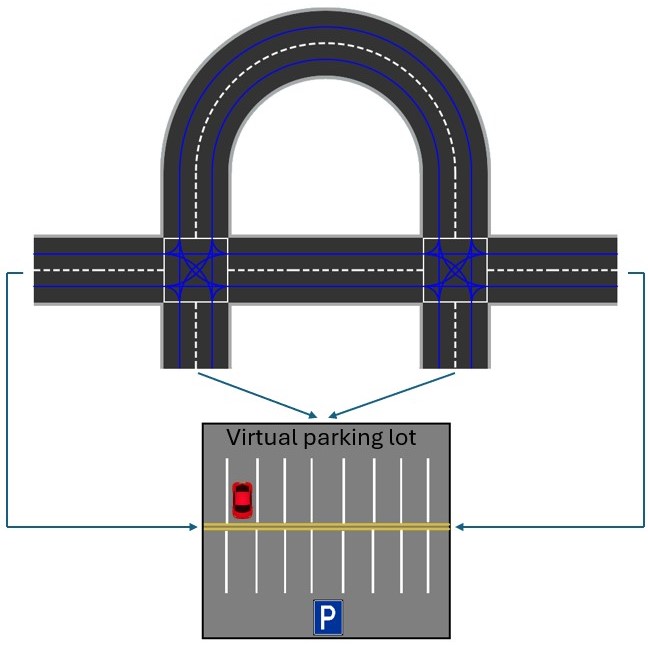}
	\caption{Illustration of the virtual parking lot handling vehicles that leave the road network through its open ends.}
	\label{fig:virtual_parking_lot}
	\vspace{-\baselineskip}
\end{figure}

\subsubsection{Graph-Based Road Network Representation} 
\begin{figure}
	\centering
	\includegraphics[width=1.0\columnwidth]{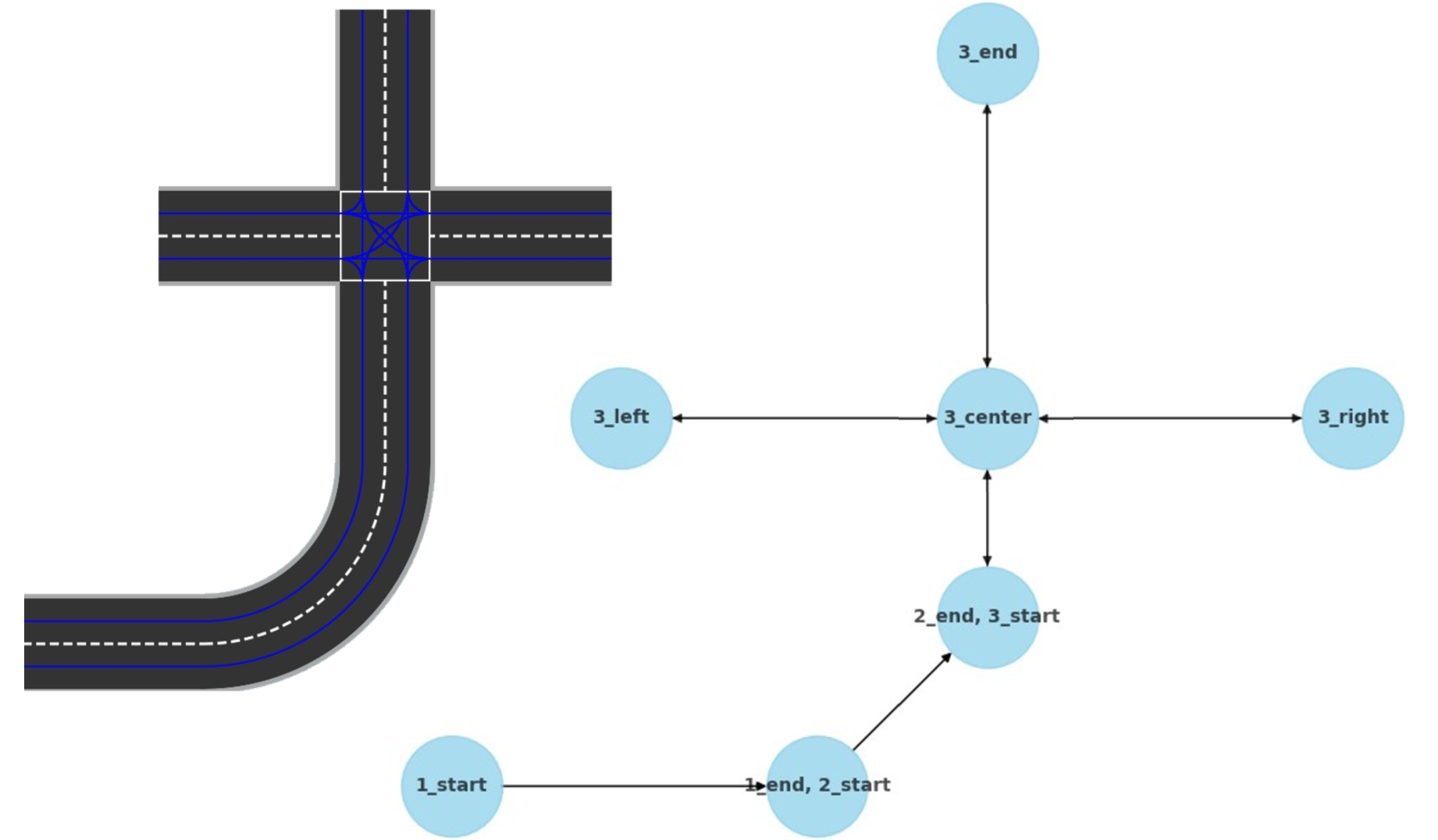}
	\caption{Road network and its graph representation.}
	\label{fig:graph_representation}
	\vspace{-\baselineskip}
\end{figure}

The simulator maintains in the \texttt{RoadSegmentManager} a graph-based representation of the road network as a \texttt{MultiDiGraph} from the NetworkX library~\cite{Hagberg2008}. Its nodes correspond to the \texttt{ConnectionPoint}s of the segments, and the edges to the \texttt{RoadSegment}s. It forms the basis for performing various of the aforementioned operations on the road network. The graph representation can be plotted with \texttt{visualize\_road\_network()} facilitating the road network construction by visualizing the already created and connected road segments including their names, see Fig.~\ref{fig:graph_representation}.

\subsection{Vehicle Manager}
\label{sec:VehicleManager}
\begin{figure}
	\centering
	\includegraphics[width=1.0\linewidth]{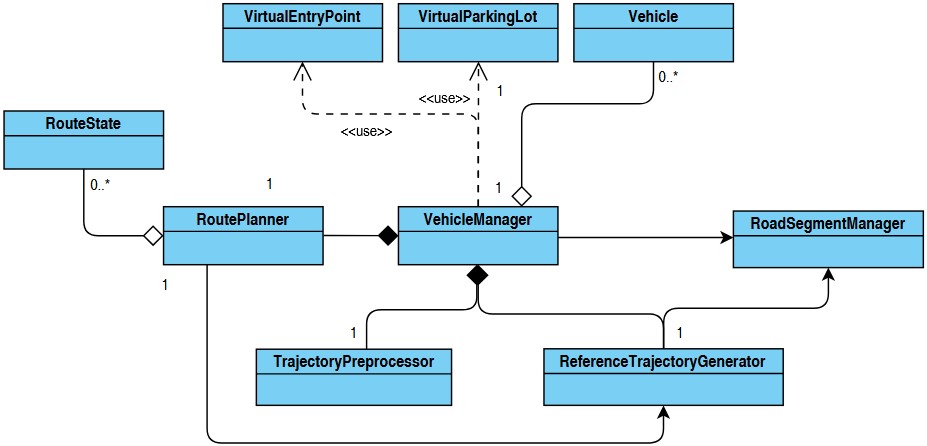}
	\caption{Class diagram of the \texttt{VehicleManager}.}
	\label{fig:VehicleManager}
\end{figure}
The simulator's vehicle system is centered around the \texttt{VehicleManager}, which manages the life cycle and coordination of all vehicles in the environment. It maintains a registry of all vehicles in \texttt{self.vehicles}, and together with its associated components it is responsible for creating, placing, updating, and tracking each vehicle throughout the simulation's runtime. The \texttt{VehicleManager} holds a reference to the \texttt{RoadSegmentManager} in its attributes, as well as to the \texttt{ReferenceTrajectoryGenerator}, the \texttt{RoutePlanner}, and the \texttt{TrajectoryPreprocessor}, see Fig.~\ref{fig:VehicleManager}. 

\subsubsection{Vehicles}
\begin{figure}
	\centering
	\includegraphics[width=1.0\linewidth]{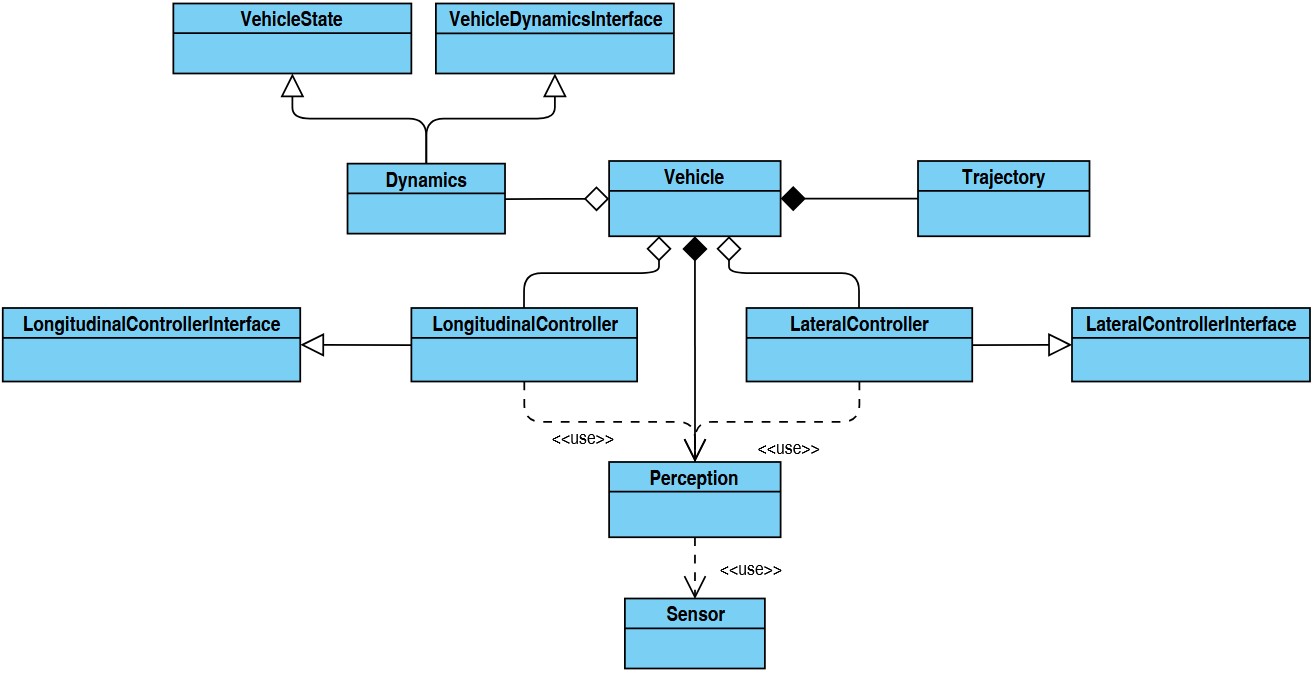}
	\caption{Class diagram of the \texttt{Vehicle}. The longitudinal and lateral controller can be alternatively replaced by a single unified controller.}
	\label{fig:vehicle_architecture}
	\vspace{-\baselineskip}
\end{figure}
\begin{figure}
	\centering
	\includegraphics[width=1.0\linewidth]{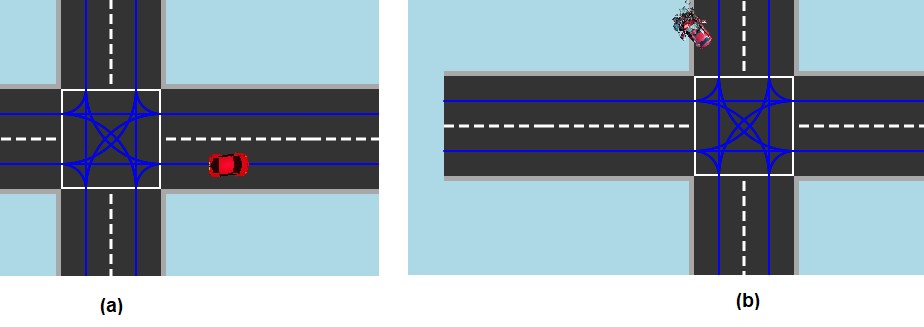}
	\caption{Visualization of \texttt{vehicle.status} in the simulation video: (a)~\texttt{"active"}, (b)~\texttt{"crashed"}.}
	\label{fig:before_after_crash}
	\vspace{-\baselineskip}
\end{figure}

During the initialization of the \texttt{VehicleManager}, vehicles are created as instances of the class \texttt{Vehicle} by calling the method \texttt{create\_vehicle()} in the manager class. A vehicle is defined in terms of its dynamics, a controller (implemented either as a \texttt{CombinedController} or split up into a \texttt{LongitudinalController} and a \texttt{LateralController}), a \texttt{Perception} module and a reference \texttt{Trajectory}. The class diagram of the \texttt{Vehicle} class is given in Fig.~\ref{fig:vehicle_architecture}. During simulation, the \texttt{VehicleManager} continuously determines if the vehicle is moving within the lane boundaries or is about to leave them, and sets the \texttt{vehicle.status} correspondingly as \texttt{"active"} or \texttt{"crashed"}; the status is visualized in the simulation video as shown in Fig.~\ref{fig:before_after_crash}. Vehicles that transition to the \texttt{VirtualParkingLot} are assigned the status \texttt{"parked"}.

\subsubsection{Vehicle Dynamics} The dynamics of a vehicle need to be defined in classes that implement the \texttt{VehicleState} interface and the \texttt{VehicleDynamicsInterface}. In this way, any dynamics can be considered that take the form
\begin{align*}
	\dot{x} = f(x,u,t), \qquad x(0) = x_{0},
\end{align*}
where $ x\in\bbR $ denotes the vehicle state, $ u\in\bbR^{m} $ the control input, $ t\in\bbR_{\geq 0} $ the simulation time, and $ x_{0} $ the vehicle's state upon initialization. The simulator exemplarily defines the kinematic bicycle model~\cite{Wang2001} in the class \texttt{BicycleDynamics}. Upon simulation, the \texttt{VehicleSimulator} simulates each vehicle over one time step~$ \Delta t $ by calling \texttt{Vehicle.update()}, solving the dynamics with \texttt{odeint()}.

\subsubsection{Perception and Sensors} The \texttt{Sensor} of a vehicle abstracts the on-board sensing and determines the relative position, relative velocity, distance and lane association of neighboring vehicles within a configurable field-of-view and range. The \texttt{Perception} module organizes the sensor output into a \texttt{PerceptionData} class, transformed into the vehicle's body frame, which is then available to the controller. An optional noise can be added to the measurements. 

\subsubsection{Reference Trajectory} 
\begin{figure}
	\centering
	\includegraphics[width=0.35\linewidth]{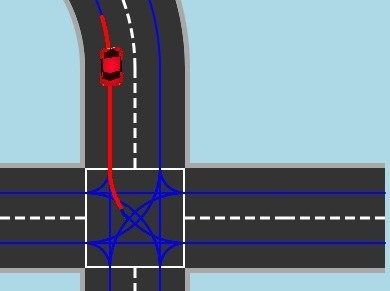}
	\caption{Visualization of lane center and reference trajectory in the simulation video.}
	\label{fig:reference_trajectory}
	\vspace{-\baselineskip}
\end{figure}

The simulator provides the possibility to generate a reference trajectory based on high-level route instructions given by a fictitious driver or a route planner. Such high-level route instructions are formulated as a list of primitives \texttt{"straight"} for following the current lane, \texttt{"left\_turn"} and \texttt{"right\_turn"} for turning to the left or the right at intersections, as well as \texttt{"left"} and \texttt{"right"} for moving to the lane to the left or right of the current lane in driving direction. The reference trajectories are based on the lane centers (marked blue in Fig.~\ref{fig:reference_trajectory}) and possibly span multiple road segments. They can be accessed through the attribute \texttt{current\_trajectory} of an instance of the class \texttt{Trajectory}. The \texttt{TrajectoryPreprocessor} allows to shorten the reference trajectory such that it only contains points lying ahead of the vehicle by calling \texttt{preprocess()}. The route instruction primitives are evaluated segment-wise (one primitive is executed per road segment). Route instructions can be updated via the \texttt{RoutePlanner}. The reference trajectory simplifies the integration of high-level route instructions into the controller implementation and renders the vehicle coordination simulation more realistic. We point out that upon lane change (primitive \texttt{"left"} or \texttt{"right"}) the reference trajectory is discontinuous and the actual lane change is executed by the low-level controllers following the route instructions while accounting for inter-vehicle coordination. Reference trajectories are given in global coordinates. It is visualized in the simulation video by a red line, see Fig.~\ref{fig:reference_trajectory}.

\subsubsection{Controller} The controller of the vehicles is implemented either as \texttt{CombinedController} computing all control inputs together, or by splitting the controller into a \texttt{LateralController} and a \texttt{LongitudinalController} computing steering angle and velocity/acceleration separately, see Fig.~\ref{fig:reference_trajectory}. The controllers can access \texttt{PerceptionData} and the reference trajectory for the control input computation. Exemplary control objectives can be lane tracking, adaptive cruise control (ACC), platoon merging and splitting, and inter-vehicle coordination upon lane change, but also advanced control objectives such as vehicle coordination at intersections, traffic-flow enhancement, leader-follower cooperation, and many more. 

\begin{figure}
	\centering
	\includegraphics[width=1.0\linewidth]{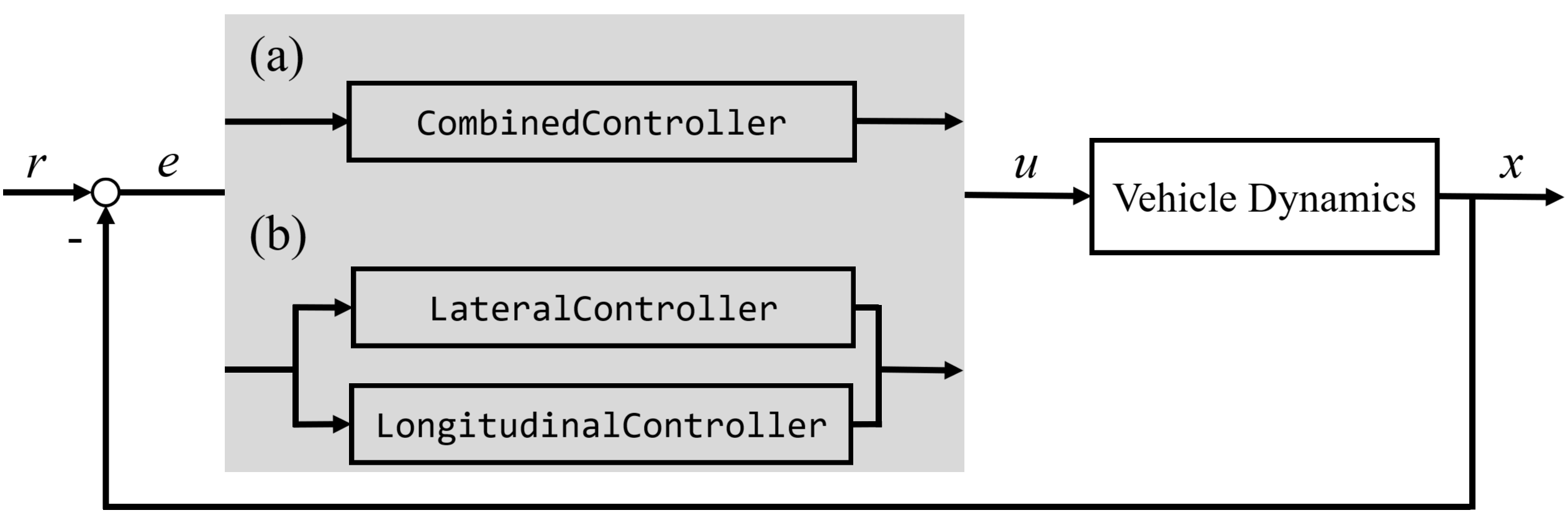}
	\caption{Closed-loop control system.}
	\vspace{-\baselineskip}
\end{figure}

\subsubsection{Simulation} At each simulation step, every vehicle executes the following update steps: first, the route instructions are updated and the reference trajectory generated; secondly, \texttt{PerceptionData} is updated; thirdly, the control inputs are computed by evaluating the controller; at last,	\label{fig:control_loop} the vehicle state is updated by integrating the vehicle dynamics over one time step~$ \Delta t $.

\subsection{Visualization} The simulation results can be visualized as a video using the \texttt{Visualization} class. To enable this feature, it is sufficient to set \texttt{SAVE\_VIDEO = True} in the \texttt{config.py} file; all other steps are handled internally by the simulator. The centering of the road network in the video is adjusted automatically. The \texttt{Visualization} class is implemented using the Pygame package. An example of the resulting visualization can be found on Youtube (see link in abstract).

\section{Using the Simulator - A Practitioner Guide}

To facilitate the setup of new simulation studies, the simulator comes along with a folder structure that can be taken as a starting point for the implementation. We suggest to proceed as follows.

\subsection{Entry point for running simulations} 
Once the simulation study has been set up, it can be executed via \texttt{main.py}, which calls the modules used to configure the simulation scenario described in the following sections.

\subsection{Setup of the Road Network} 
Navigate to \texttt{toolbox/initialization/road\_} \texttt{setup.py} and define the necessary road segments using the \texttt{create\_road\_segment()} method from \texttt{Traffic}\texttt{Environment}. For example, 
\begin{lstlisting}[language=Python]
environment.create_road_segment(
  segment_type=1, length=100.0, 
  orientation=0.0, lane_width=50, 
  lanes=2, speed_limit = 50)
\end{lstlisting}
Here, \texttt{segment\_type = 1} represents a straight road, while 2 and 3 correspond to a curved road and an intersection, respectively. The exact arguments for \texttt{create\_road\_segment()} may vary slightly depending on the selected \texttt{segment\_type}. For assembling the created road segments to a road network, navigate to \texttt{connection\_setup.py}. Use the \texttt{connect\_road\_segments()} method to link the segments together. For example, to connect the \texttt{"start"} of segment~2 to the \texttt{"end"} of segment~1, call
\begin{lstlisting}[language=Python]
environment.connect_road_segments(
  fixed_segment_index=1, 
  connection_point_1="end", 
  moving_segment_index=2, 
  connection_point_2="start")
\end{lstlisting}
The segment designated as the moving segment is thereby automatically aligned with the fixed segment by adjusting its position and orientation accordingly. To connect open ends in the resulting road network, navigate to \texttt{open\_end\_automatic\_connection.py} and call
\begin{lstlisting}[language=Python]
environment.automatically_generate_road_
  to_connect_open_end_segments(...)
\end{lstlisting}
Finally, configure the virtual parking lot in \texttt{virtual\_} \texttt{parking\_lot\_setup.py}, for example, as follows:
\begin{lstlisting}[language=Python]
environment.create_virtual_parking_lot(
  platoon_size=2,
  exit_points=[
  {'segment_id': "2_start", 
   'connection_point': 'start'},],
  time_sequence_interval=4, 
  time_mean=5, time_variance=0)
\end{lstlisting}
This configuration defines the platoon size of vehicles re-entering the road network, the re-entry locations specified in \texttt{exit\_points}, the time interval between consecutive vehicle releases within a platoon, and the mean and variance governing the delay before a completed platoon is reintroduced into the network.

\subsection{Vehicle Creation and Controller Implementation}

Begin by implementing the controller either as a unified controller using the \texttt{CombinedController} interface, or as separate lateral and longitudinal components using the \texttt{LateralController} and \texttt{LongitudinalController} interfaces, respectively. These can be defined in \texttt{toolbox/vehicles/} \texttt{controllers.py}, which also contains example implementations based on the bicycle kinematic model. Similarly, vehicle dynamics can be implemented in \texttt{dynamics.py} by defining a class that implements the \texttt{VehicleDynamicsInterface}. A kinematic bicycle model is provided as a reference implementation. Based on these components, individual vehicles are defined in \texttt{toolbox/initialization/vehicle\_creation.py} via the following method
\begin{lstlisting}[language=Python]
def create_vehicle(
  self, 
  vehicle_id: int, 
  dynamic_model: VehicleDynamicsInterface, 
  controller: CombinedController = None,  
    lateral_controller: 
  LateralController = None, 
  longitudinal_controller: 
    LongitudinalController = None)
\end{lstlisting}
Vehicles can be placed in \texttt{vehicle\_placement.py} by calling the method \texttt{add\_vehicle\_to\_segment()}.

\subsection{General Configurations}

For general simulation settings, navigate to \texttt{toolbox/} \texttt{config/config.py}. Here, global configuration variables can be adjusted to define the simulation \texttt{TIME\_STEP} and the \texttt{SIMULATION\_DURATION}, as well as to enable the generation of simulation videos. In \texttt{vehicle\_data\_logging.py}, it can be configured which data is logged during the simulation for later analysis. For example the vehicle positions, velocities or control inputs can be logged over specified time intervals and sampling rates.

\section{Getting Started - Exemplary Simulation Studies}

\begin{figure}
	\centering
	
	\begin{subfigure}[b]{0.37\linewidth}
		\centering
		\includegraphics[width=\linewidth]{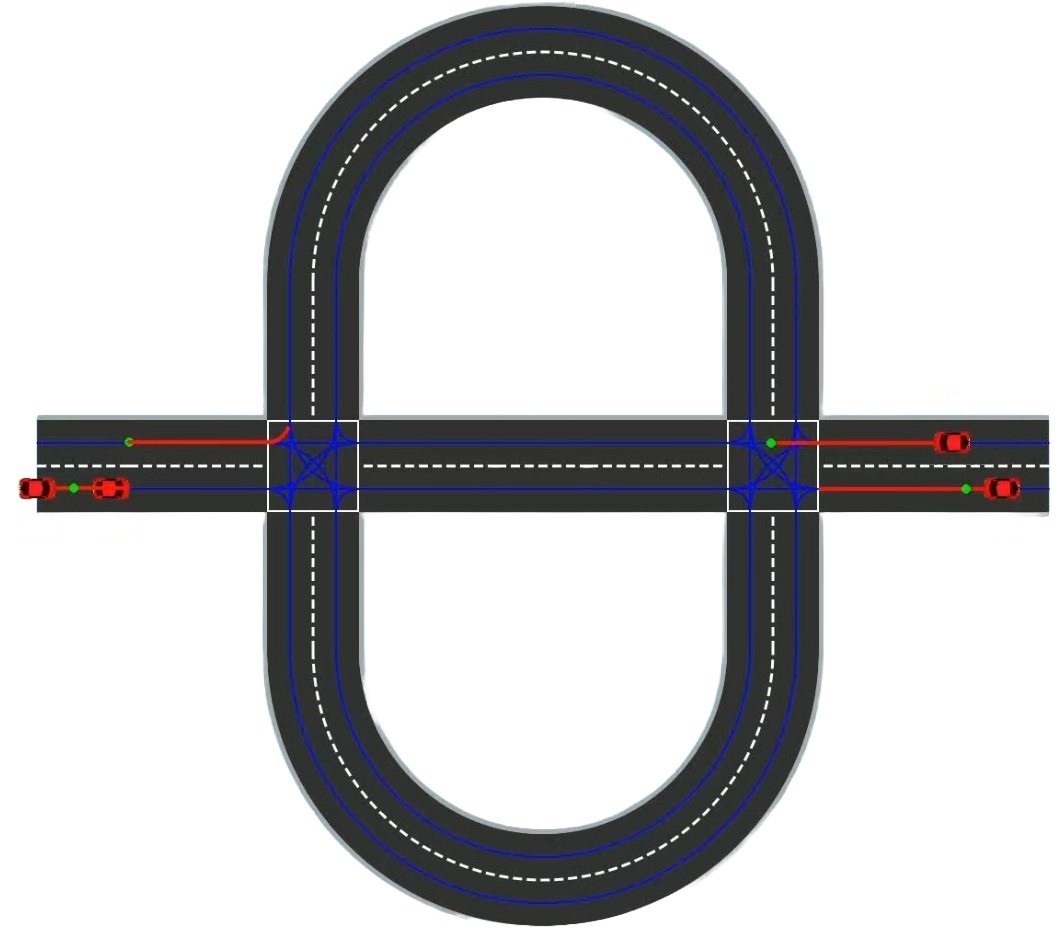}
		\label{fig:vehicle-placements}
	\end{subfigure}
	\hfill
	\begin{subfigure}[b]{0.6\linewidth}
		\centering
		\includegraphics[width=\linewidth]{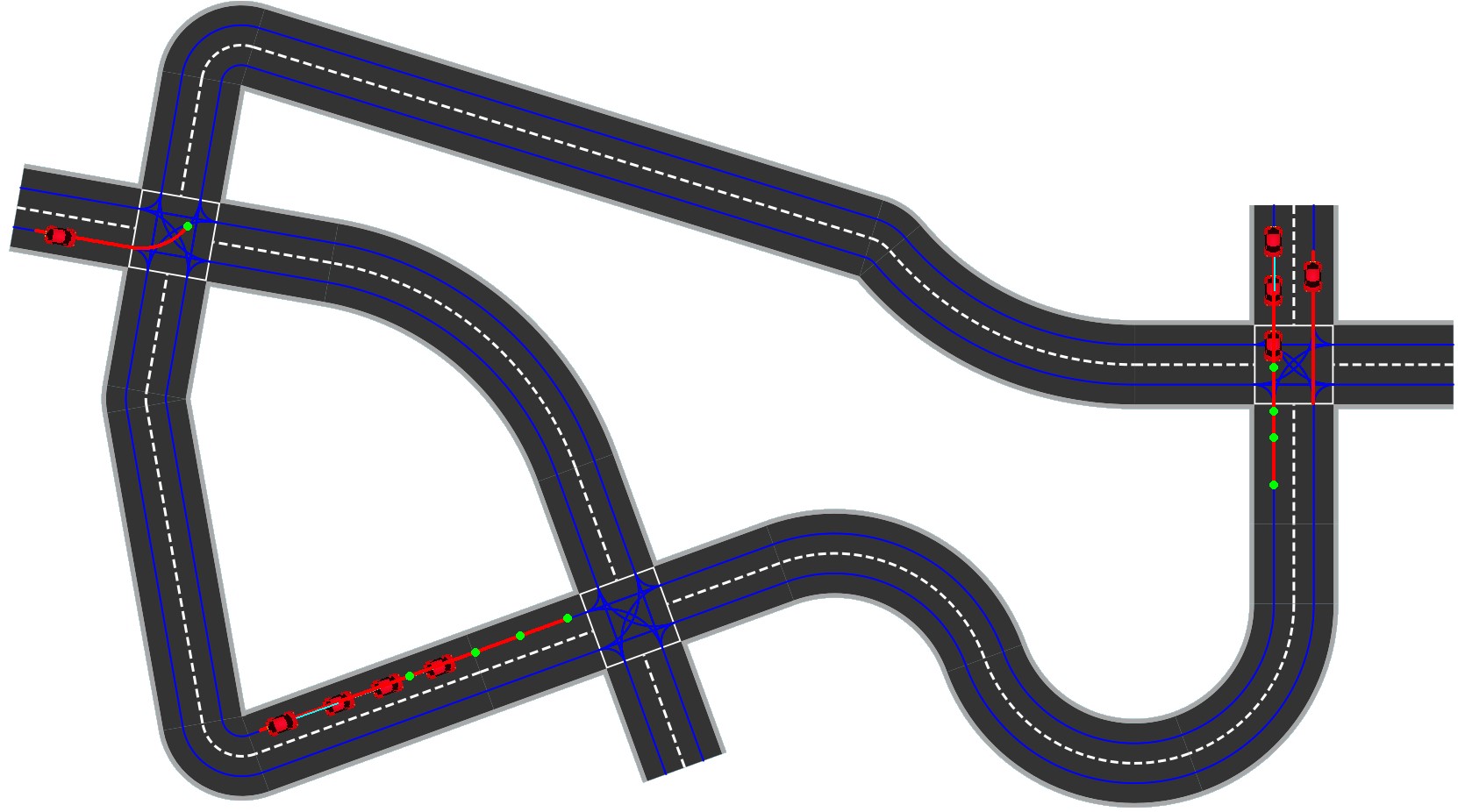}
		\label{fig:complex-vehicle-placements}
	\end{subfigure}
	\caption{Getting started -- two distinct exemplary simulation studies are provided along with the simulator framework.}
	\label{fig:vehicle-placement-comparison}
\end{figure}

The simulator framework comes along with two simulation studies on different road networks with increasing complexity, which can be used as a starting point for implementing new simulation scenarios. The implemented controllers account for lane tracking, adaptive cruise control, lane changing as well as the merging and splitting of platoons. Screenshots from the simulation are presented in Fig.~\ref{fig:vehicle-placement-comparison} showing the two different considered road networks. By running \texttt{main.py} in the initial configuration starts the simulation of the second simulation scenario. The saved simulation results, including videos, can be found in the folder \texttt{simulation\_projects}.

\section{Discussion \& Conclusion}
\label{sec:conclusion}

The presented simulator is designed to support research and education on platooning, vehicle coordination, and  (distributed) control. By providing a unified simulation framework, it lowers the implementation overhead for novel control and coordination algorithms in complex traffic scenarios and to visualize the simulation results.
Owing to its modular python implementation, it is both straightforward to use and highly customizable, allowing for various extensions. In particular, the framework can be expanded to incorporate explicit inter-vehicle communication, integration with IoT infrastructure, and human interaction, thereby enabling interactive studies in mixed traffic environments involving both autonomous and human-operated vehicles.
At an educational level, students can implement their own control algorithms and directly receive visual feedback by running their solution in a traffic environment pre-configured by the instructor.


\bibliographystyle{IEEEtran}

\balance

\bibliography{/home/adrian-wiltz/MyCloudStation/JabBib/Research/000_MyLibrary}

\begin{thebibliography}{10}
\providecommand{\url}[1]{#1}
\csname url@samestyle\endcsname
\providecommand{\newblock}{\relax}
\providecommand{\bibinfo}[2]{#2}
\providecommand{\BIBentrySTDinterwordspacing}{\spaceskip=0pt\relax}
\providecommand{\BIBentryALTinterwordstretchfactor}{4}
\providecommand{\BIBentryALTinterwordspacing}{\spaceskip=\fontdimen2\font plus
\BIBentryALTinterwordstretchfactor\fontdimen3\font minus
  \fontdimen4\font\relax}
\providecommand{\BIBforeignlanguage}[2]{{%
\expandafter\ifx\csname l@#1\endcsname\relax
\typeout{** WARNING: IEEEtran.bst: No hyphenation pattern has been}%
\typeout{** loaded for the language `#1'. Using the pattern for}%
\typeout{** the default language instead.}%
\else
\language=\csname l@#1\endcsname
\fi
#2}}
\providecommand{\BIBdecl}{\relax}
\BIBdecl

\bibitem{Botelho2025}
T.~C. Botelho, S.~P. Duarte, M.~C. Ferreira, S.~Ferreira, and A.~Lobo,
  ``Simulator and on-road testing of truck platooning: a systematic review,''
  \emph{European Transport Research Review}, vol.~17, no.~1, p.~4, 2025.

\bibitem{Dosovitskiy2017}
A.~Dosovitskiy, G.~Ros, F.~Codevilla, A.~Lopez, and V.~Koltun, ``{CARLA}: {An}
  open urban driving simulator,'' in \emph{Proceedings of the 1st Annual
  Conference on Robot Learning}, 2017, pp. 1--16.

\bibitem{Lopez2018}
P.~A. Lopez, M.~Behrisch, L.~Bieker-Walz, J.~Erdmann, Y.-P. Flötteröd,
  R.~Hilbrich, L.~Lücken, J.~Rummel, P.~Wagner, and E.~Wiessner, ``Microscopic
  traffic simulation using sumo,'' in \emph{2018 21st International Conference
  on Intelligent Transportation Systems (ITSC)}, 2018, pp. 2575--2582.

\bibitem{Althoff2017}
M.~Althoff, M.~Koschi, and S.~Manzinger, ``Commonroad: Composable benchmarks
  for motion planning on roads,'' in \emph{2017 IEEE Intelligent Vehicles
  Symposium (IV)}, 2017, pp. 719--726.

\bibitem{Frauenfelder2023}
A.~Frauenfelder, A.~Wiltz, and D.~V. Dimarogonas, ``Decentralized vehicle
  coordination and lane switching without switching of controllers,''
  \emph{IFAC-PapersOnLine}, vol.~56, no.~2, pp. 3334--3339, 2023.

\bibitem{Charitidou2022b}
M.~Charitidou and D.~V. Dimarogonas, ``Splitting and merging control of
  multiple platoons with signal temporal logic,'' in \emph{2022 IEEE Conference
  on Control Technology and Applications (CCTA)}, 2022, pp. 1031--1036.

\bibitem{omg2017}
\BIBentryALTinterwordspacing
{Object Management Group}, \emph{OMG Unified Modeling Language}, 2017.
  [Online]. Available: \url{https://www.omg.org/spec/UML/2.5.1/PDF}
\BIBentrySTDinterwordspacing

\bibitem{Fowler2004}
M.~Fowler, \emph{\BIBforeignlanguage{eng}{UML distilled : a brief guide to the
  standard object modeling language}}, 3rd~ed., ser. Addison-Wesley object
  technology series.\hskip 1em plus 0.5em minus 0.4em\relax Boston, Mass:
  Addison-Wesley, 2004.

\bibitem{Dubins1957}
L.~E. Dubins, ``On curves of minimal length with a constraint on average
  curvature, and with prescribed initial and terminal positions and tangents,''
  \emph{American Journal of Mathematics}, vol.~79, no.~3, pp. 497--516, 1957.

\bibitem{Jamsahar2025}
K.~Jamsahar, ``Design of a modular platooning and traffic coordination
  simulator,'' Master's thesis, KTH Royal Institute of Technology, 2025,
  supervised by Adrian Wiltz and Maria Charitidou.

\bibitem{Hagberg2008}
A.~Hagberg, P.~J. Swart, and D.~A. Schult, ``Exploring network structure,
  dynamics, and function using networkx.''\hskip 1em plus 0.5em minus
  0.4em\relax Los Alamos National Laboratory (LANL), 01 2008.

\bibitem{Wang2001}
D.~Wang and F.~Qi, ``Trajectory planning for a four-wheel-steering vehicle,''
  in \emph{Proceedings 2001 ICRA. IEEE International Conference on Robotics and
  Automation}, vol.~4, 2001, pp. 3320--3325.

\end{thebibliography}


\end{document}